\documentclass[aps,prl,reprint]{revtex4-1}
\usepackage{amsmath}
\usepackage{graphicx}
\usepackage{color}
\usepackage{amssymb}
\usepackage{wasysym}
\usepackage{amsthm}
\usepackage{soul}
\usepackage{mathtools}   

\input{epsf}

\begin{document}

\newcommand{\rvec}{\textbf{r}}
\newcommand{\drvec}{\Delta\textbf{r}}
\newcommand{\omgvechat}{\hat{\boldsymbol{\omega}}}
\newcommand{\chivechat}{\hat{\boldsymbol{\chi}}}
\newcommand{\nvechat}{\hat{\mathbf{n}}}
\newcommand{\legpol}{\mathcal{P}}
\newcommand{\rrot}{\mathcal{R}}
\newcommand{\fe}{\mathcal{F}}
\newcommand{\feid}{\mathcal{F}_{\rm id}}
\newcommand{\fex}{\mathcal{F}_{\rm exc}}
\newcommand{\intleg}{\int_{-1}^1}
\newcommand{\qmin}{q_{\rm min}}

\newcommand*{\bfrac}[2]{\genfrac{}{}{0pt}{}{#1}{#2}}

\bibliographystyle{prsty}

\title{Density Functional Theory for Chiral Nematic Liquid Crystals}

\author{S. Belli$^1$, S. Dussi$^2$, M. Dijkstra$^2$ and R. van Roij$^1$}
\affiliation{$^1$Institute for Theoretical Physics, Utrecht University, Leuvenlaan 4, 3584 CE Utrecht, The Netherlands \\
$^2$Soft Condensed Matter, Debye Institute for Nanomaterials Science, Utrecht University, Princetonplein 5, 3584 CC Utrecht, The Netherlands}

\begin{abstract}

Even though chiral nematic phases were the first liquid crystals experimentally observed more than a century ago, the origin of the thermodynamic stability of cholesteric states is still unclear. In this Letter we address the problem by means of a novel density functional theory for the equilibrium pitch of chiral particles. When applied to right-handed hard helices, our theory predicts an entropy-driven cholesteric phase, which can be either right- or left-handed, depending not only on the particle shape but also on the thermodynamic state. We explain the origin of the chiral ordering as an interplay between local nematic alignment and excluded-volume differences between left- and right-handed particle pairs.

\medskip

\noindent PACS numbers: 64.70.mf, 64.70.pv, 61.30.Cz, 61.30.St

\end{abstract}

\maketitle

\begin{figure}[t]
\center
\includegraphics[scale=0.5]{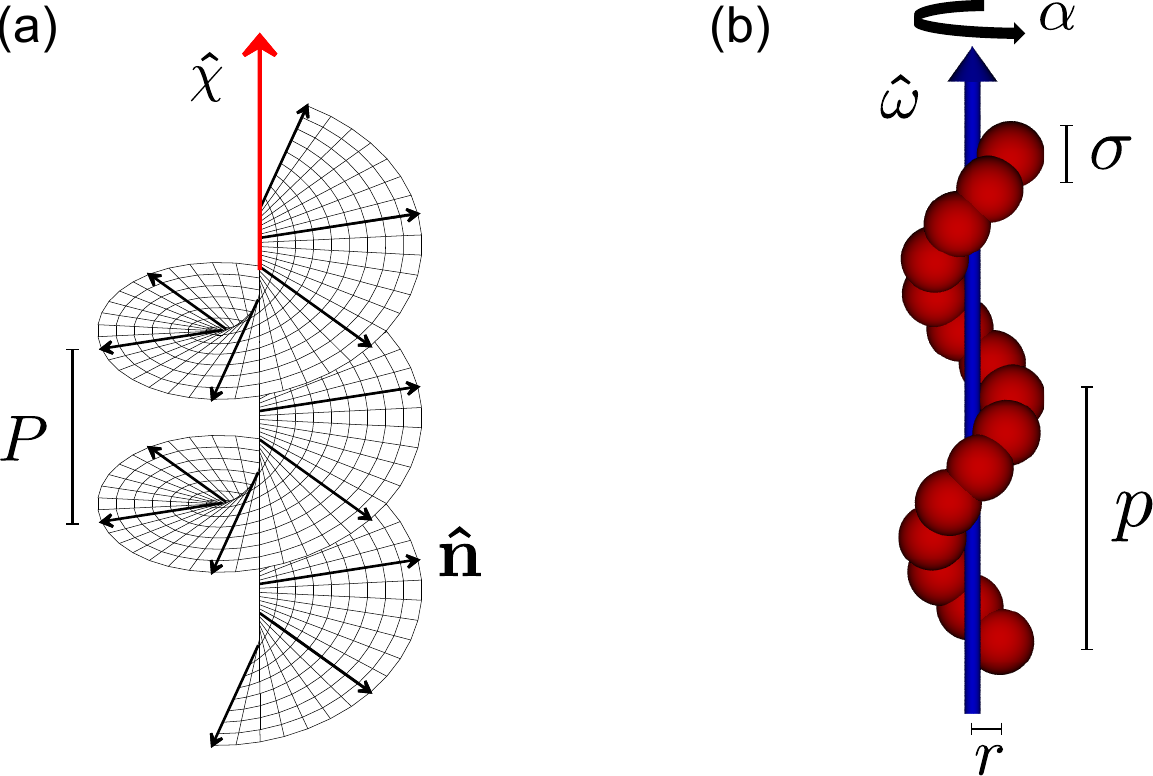}
\caption{\label{fig1} (a) Space dependence of the nematic director $\nvechat$ in a cholesteric phase with chiral director $\chivechat$ and pitch $P$. (b) Hard helix particle modeled as a collection of $15$ partially overlapping hard spheres of diameter $\sigma$, whose centers of mass are equally spaced on a right-handed helix of contour length $L=10 \, \sigma$, inner radius $r=0.4 \, \sigma$ and pitch $p=4 \,\sigma$ \cite{frezza}. The orientation of the helix can be expressed by means of the unit vector $\omgvechat$ and the internal azimuthal angle $\alpha$.} 
\end{figure}

{\em Cholesteric} phases, also known as {\em chiral nematics}, are fascinating examples of liquid crystals. Liquid crystals are phases of matter characterized by a degree of spontaneous breaking of the rotational and translational symmetries that is higher than in the liquid and lower than in the crystal phase. In the nematic phase, for example, the particles self-organize by all aligning along a common direction (the {\em nematic director}), while keeping their centers of mass homogeneously distributed in space. Chiral nematic liquid crystals are peculiar as their nematic director rotates like a helix around a {\em chiral director}, thus giving rise to a chiral distribution of the orientations of the particles \cite{degennes}. A cholesteric phase can be either right-handed (as in Fig. \ref{fig1}(a)) or left-handed, depending on the handedness of the helix drawn by the nematic director $\nvechat$. The wavelength associated to a full rotation of the nematic director around the chiral director $\chivechat$ is known as the cholesteric {\em pitch} $P$. 
Cholesteric phases are commonly found in both thermotropic molecular compounds (e.g., derivatives of cholesterol \cite{reinitzer,dupre,huff,rossi}) and in lyotropic colloidal suspensions of, e.g., DNA \cite{livolant, zanchetta} and filamentous viruses \cite{dogic,grelet,tombolato2006,barry,zhang}. Their widespread occurrence explains why cholesterics were the first liquid-crystal phase experimentally observed \cite{reinitzer}. The pitch is experimentally known to take values several orders of magnitude higher than the size of the constituent particles, reaching the visible-light wavelength in molecular compounds. For this reason, and for their liquid-like rheological properties, cholesterics have long found wide technological application in the opto-electronic industry \cite{collings}.

Despite their long history and widespread technological applications, surprisingly little is understood about this chiral state of matter. A fundamental open question regards the relationship between macroscopic and microscopic chirality, i.e., between the handedness of the phase and that of the constituent particles \cite{harris1997, harris1999, wensink2009,varga2011}. Moreover, it is still to be unambiguously proved that hard chiral interactions alone can give rise to an entropy-stabilized chiral-nematic phase \cite{varga2011, frezza}. The problem in interpreting experimental data is largely due to the limitations of theory and simulation methods. Computer simulation studies on cholesterics, for instance, require sophisticated techniques and huge simulation boxes in order to reproduce the pitch \cite{allen,frenkel}, thereby imposing severe limitations on the complexity of the inter-particle model potentials. Therefore, simulation work has focused mostly on coarse-grained potentials  \cite{memmer, germano,varga2006}, where the microscopic chirality of the pair interaction is implicitly averaged into a single pseudo-scalar parameter \cite{goossens}. As a result, the crucial question regarding the relation between microscopic and macroscopic chirality has remained unanswered as of today. In order to overcome the severe requirements of simulation, more sophisticated model potentials \cite{tombolato2005,wensink2011}, in which the chirality is introduced in full detail at the microscopic level, have been recently studied by means of Straley's classic theory for the prediction of the cholesteric pitch \cite{straley}. Despite the undoubted relevance of Straley's pioneering work, his approach suffers from two major drawbacks that limit the reliability of its predictions. First, the theory is based on a second-order small $q$ Taylor expansion of the free-energy functional, and is therefore valid only in the limit of a very long pitch $P=2 \pi/q$. Even though the latter condition holds in most experimental situations, this approximation limits the applicability of the theory to more general instances. Second, Straley's theory cannot be solved self-consistently, in the sense that the orientation distribution function in the presence of a chiral twist is assumed to be the same as in the achiral limit, an approximation of which the quality is not easily assessed.

In this Letter we develop a novel density functional theory (DFT) that overcomes these two main drawbacks of Straley's theory. Following Onsager \cite{onsager}, the interactions are introduced by truncating the virial expansion at second order in the density. Improving over Straley's theory, our method allows to numerically minimize the free-energy functional exactly (i.e., at arbitrary precision) at this virial order. Moreover, the numerical method underlying our calculations allows for the study of arbitrary pair potentials. Here we focus on the chiral nematic phase developed by a large class of hard-sphere helices described in Ref. \cite{frezza} (cf. Fig. \ref{fig1}(b)). 

The long-range orientational order of a homogeneous phase is described in terms of the orientation distribution function (ODF) $\psi(\rrot)$, which is the probability density of a particle having an orientation $\rrot$. The $3 \times 3$ rotation matrix $\rrot$ can be parameterized in terms of the unit vector $\omgvechat=(\cos \phi \sin \theta,\sin \phi \sin \theta , \cos \theta)$, where $\phi \in [0, 2 \pi)$ and $\theta \in [0, \pi)$, and the internal azimuthal angle $\alpha \in [0, 2 \pi)$ (cf. Fig. \ref{fig1}(b)). The ODF satisfies the normalization condition $ \oint d \rrot \, \psi(\rrot) = 1$, where $d \rrot = d \omgvechat \, d \alpha = d \cos \theta \, d \phi \, d \alpha$. A system of helices is in an (achiral) uniaxial nematic phase if the ODF satisfies $\psi(\rrot)=\psi(\nvechat_0 \cdot \omgvechat)$, where the nematic director $\nvechat_0$ is a spatial constant that can be chosen to be the $z$ axis of the laboratory reference frame, so that $\nvechat_0 \cdot \omgvechat=\cos \theta$. In this situation it is possible to expand the single particle density in Legendre polynomials \cite{vroege}.
Let us now consider a cholesteric phase, whose chiral director coincides, say, with the $y$ axis of the laboratory frame. Let $P$ be the macroscopic pitch and $q=2\pi/P$ the corresponding chiral wavenumber. At any point $\rvec=(x,y,z)$ in space the nematic director is $\nvechat_q(y) = \rrot_y(q y) \, \nvechat_0$,
where $\rrot_y(q y)$ is a $3 \times 3$ matrix representing a rotation around the $y$ axis by an angle $q y$ and $\nvechat_0=\nvechat_q(0)$. The ODF describing a chiral-nematic phase can then be expanded in Legendre polynomials $\legpol_l(x)$ as 

\begin{equation}
\psi \bigl(\nvechat_q(y) \cdot \omgvechat \bigr) =  \sum_{l=0}^\infty \psi_l \, \legpol_l \bigl(\nvechat_q(y) \cdot \omgvechat \bigr),
\label{eq4}
\end{equation}
with the expansion coefficients $\psi_l$ given by

\begin{equation}
\psi_l = \frac{2l+1}{2} \int_{-1}^1 d(\nvechat_0 \cdot \omgvechat)\, \psi(\nvechat_0 \cdot \omgvechat)\, \legpol_l (\nvechat_0 \cdot \omgvechat), 
\label{eq2}
\end{equation}
where $d(\nvechat_0 \cdot \omgvechat)=d\cos \theta$. Note that $\psi_l$ does {\em not} depend on $q$. The Legendre expansion in Eq. (\ref{eq4}) is particularly convenient since it decouples the information on the local distribution of orientations (i.e., the set of coefficients $\psi_l$) from that about the chiral period $q$. In other words, once the ODF at $y=0$ and the pitch $P$ are separately known, we can reconstruct the ODF at arbitrary $y$ by means of Eq. (\ref{eq4}) and (\ref{eq2}).

We calculate the equilibrium ODF $\psi \bigl(\nvechat_q(y) \cdot \omgvechat \bigr)$ of hard helices by means of DFT \cite{evans}. Within DFT the free energy is expressed as a sum of an ideal-gas and an excess functional, $\fe[\psi]=\feid[\psi]+\fex[\psi]$. The ideal component of the free-energy functional per unit volume is known exactly and consists of a translational and rotational entropic contribution given by

\begin{multline}
\frac{\beta \feid[\psi]}{V} =  \rho \bigl [ \log (\rho \Lambda^3) - 1 \bigr ] + \\
+ 4 \pi^2 \rho \intleg d (\nvechat_0 \cdot \omgvechat)\, \psi \bigl(\nvechat_0 \cdot \omgvechat \bigr) \log\bigl[ \psi \bigl(\nvechat_0 \cdot \omgvechat \bigr) \bigr],
\label{eq5}
\end{multline}
where $\beta=(k_B T)^{-1}$, $k_B$ is the Boltzmann constant, $T$ the temperature, $\Lambda^3$ the thermal volume and $\rho=N/V$ the number density \cite{onsager, vroege}. The absence of any dependence on $q$ in Eq. (\ref{eq5}) renders explicit the fact that the ideal-part of the free energy is insensitive to chiral ordering. The excess free-energy functional, which accounts for particle-particle interactions, is not known and has thus to be approximated. Here we adopt the Parsons-Lee second-virial approximation, which is known to be exact for infinitely thin rods in the achiral limit $q = 0$ \cite{onsager} and interpolates smoothly via shorter rods to spheres \cite{plee, bolhuis}. By introducing the Legendre polynomial expansion Eq. (\ref{eq4}), the second-virial excess free-energy functional can be expressed as

\begin{equation}
\frac{\beta \fex[\psi]}{V} = \frac{\rho^2 \, G(\rho v)}{2} \sum_{l,l'=0}^{\infty} \psi_l \, \psi_{l'} \, E_{l l'}(q),
\label{eq6}
\end{equation}
where $v$ is the single-particle volume and $G(x)=\frac{1-3x/4}{(1-x)^2}$ is the Parsons-Lee correction term \cite{plee}.
The $q$-dependent coefficients $E_{l l'}(q)$ are defined as

\begin{multline}
E_{l l'}(q) = - \int d (\drvec) \oint d \rrot \, d \rrot' \times \\
\times f \bigl(\drvec, \rrot, \rrot' \bigr)\, \legpol_l \bigl(\nvechat_0 \cdot \omgvechat \bigr) \, \legpol_{l'} \bigl( \nvechat_q(\Delta y) \cdot \omgvechat' \bigr).
\label{eq7}
\end{multline}
The Mayer function $f = e^{-\beta u} -1$ is defined in terms of the pair potential $u(\drvec, \rrot, \rrot')$ of two particles with orientations $\rrot$ and $\rrot'$, respectively, and center-to-center distance $\drvec=\rvec-\rvec'$. Notice that, by setting $q=0$ in Eq. (\ref{eq7}), we recover the usual Legendre polynomial expansion of the excluded volume \cite{vroege}.

\begin{figure}
\includegraphics[scale=0.7]{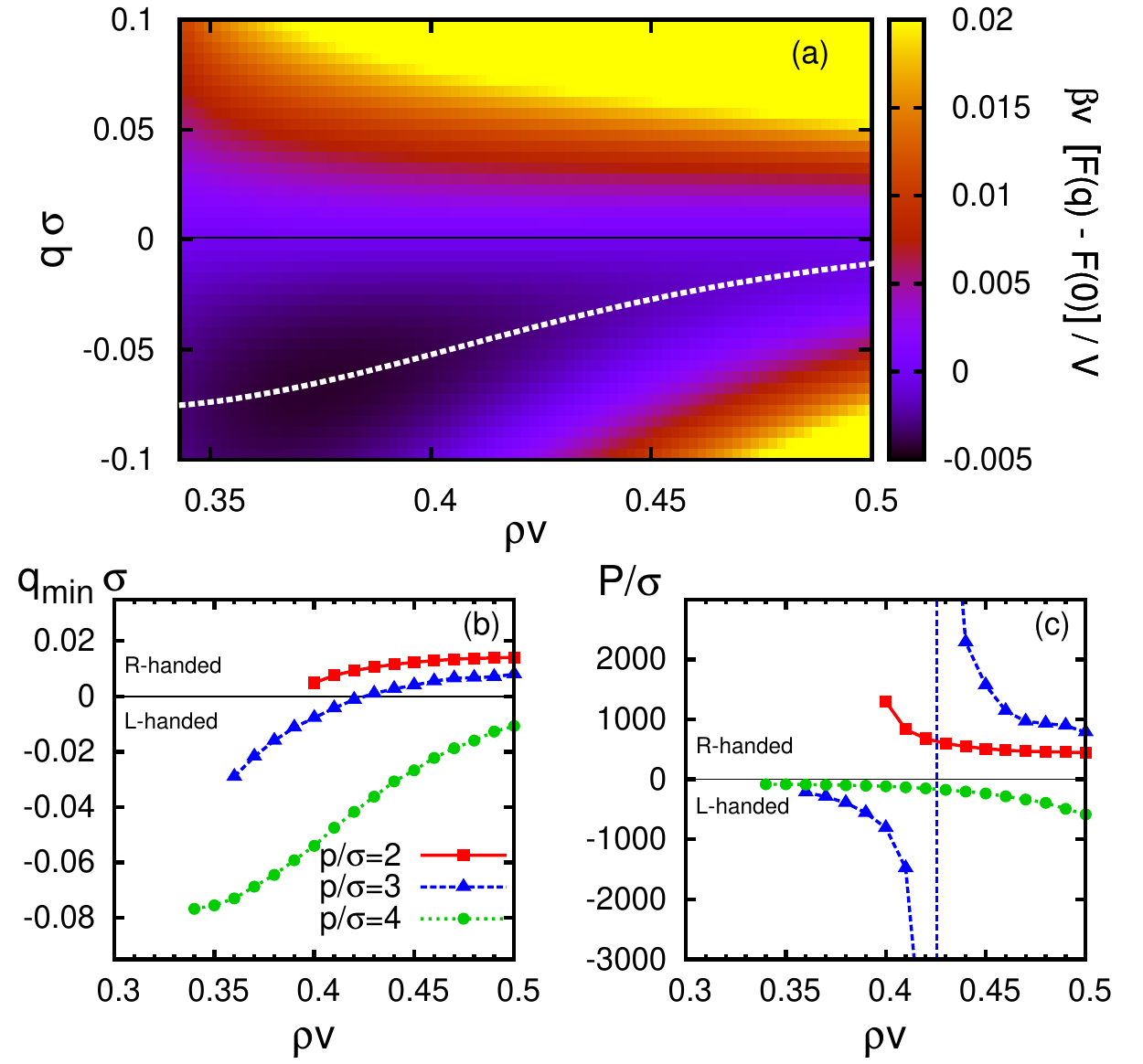}
\caption{\label{fig2} (a) Free energy per unit volume as a function of the packing fraction $\rho v$ and the chiral wavenumber $q$ for hard helices with contour length $L/\sigma=10$, helix radius $r/\sigma=0.4$ and pitch $p/\sigma=4$. The white dashed line identifies $q_{\rm min}$ that minimizes the free energy at fixed packing fraction. (b) Chiral wavenumber $q_{\rm min}$ for hard helices with contour length $L/\sigma=10$, helix radius $r/\sigma=0.4$ and helix pitch $p/\sigma=2$, $3$ and $4$ and (c) the corresponding equilibrium pitch $P=2 \pi/\qmin$.} 
\end{figure}

The advantage of expressing the excess free-energy functional in terms of the coefficients $E_{l l'}(q)$ as in Eq. (\ref{eq6}) is immediately evident. By separating the information about the local ordering of the particles (the Legendre coefficients $\psi_l$) from that about the pitch (the chiral wavenumber $q$), we can minimize the functional in three standard steps. First, we evaluate via Monte Carlo integration the coefficients $E_{l l'}(q)$ in Eq. (\ref{eq7}) for different values of $q$ and $0 \le l,l' \le l_{\rm max}$; in all the cases studied here the truncation $l_{\rm max}=20$ is large enough to be effectively infinity. Note that this integration needs to be done only once for a given particle shape. Second, at fixed $T$, $\rho$ {\em and} $q$ we minimize the total free energy with respect to the coefficients $\psi_l$. Finally, we reinsert the coefficients $\psi_l$ into the free energy and identify the chiral wavenumber $q_{\rm min}$ that minimizes the free energy with respect to $q$, at fixed $\rho$ and $T$. 

\begin{figure}[b]
\includegraphics[scale=0.65]{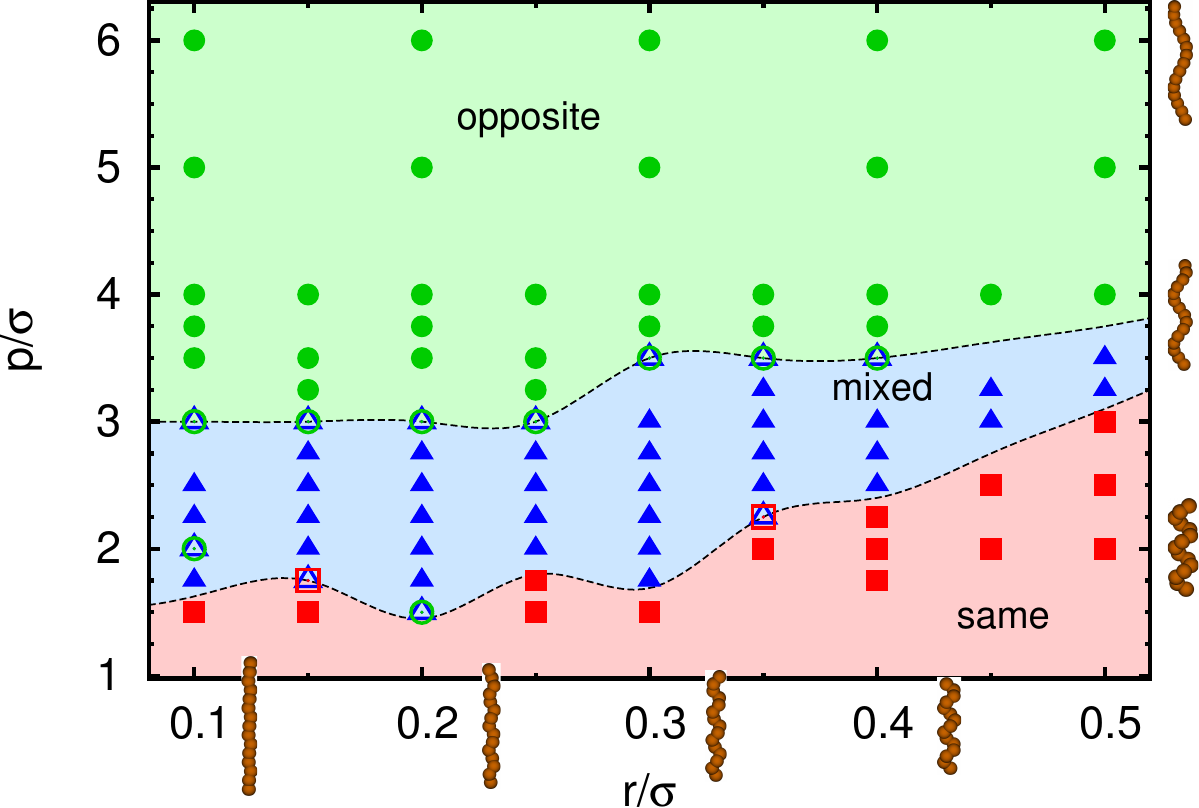}
\caption{\label{fig3} Handedness of the cholesteric phase of hard helices with contour length $L/\sigma=10$ and variable helix radius $r$ and pitch $p$. The handedness of the cholesteric phase with respect to that of the constituent hard helices can be the same ($\blacksquare$), opposite ($\CIRCLE$), or mixed with a change in handedness from left to right upon increasing the packing fraction ($\blacktriangle$). The dashed lines identify the approximate transition between these regimes. Open symbols represent parameters for which such a classification is uncertain within our statistical accuracy.}
\end{figure}

Here we apply our theory to right-handed hard helices. Fig. \ref{fig2} shows results for the shape parameters $L/\sigma=10$, inner radius $r/\sigma=0.4$, and internal pitch $p/\sigma=2$, $3$, and $4$ (see Fig.1(b)), for which the particle volume $v/\sigma^3=6.85, 6.96, 7.00$, respectively. Fig. \ref{fig2}(a) shows the $q$-dependent free-energy landscape for $p=4\, \sigma$, the white dashed line indicating $\qmin$. Fig. \ref{fig2}(b) and (c) show the dependence of $\qmin$ and the corresponding equilibrium pitch $P$ on the density, respectively. A positive value of the chiral director $\qmin$ (and pitch $P$) represents a right-handed chiral phase, whereas a negative one identifies a left-handed chiral phase. Fig. \ref{fig2}(a) shows that the free energy for $q=\qmin$ is lower than for $q=0$, and Figs. \ref{fig2}(b) and (c) demonstrate that the sole internal chirality of the particles does {\em not} determine the chirality of the phase. In fact, even though all the three helix models have the same right-handed chirality, the phases formed by helices with $p/\sigma=2$ and $p/\sigma=4$ have opposite handedness. Interestingly, helices with $p/\sigma=3$ give rise to a left-handed phase at low density and a right-handed phase at higher density. The transition between these two regimes occurs at a packing fraction $\rho v \approx 0.43$, where the phase becomes achiral ($\qmin=0$ and $P \rightarrow \infty$). An extensive investigation of the handedness of the stable cholesteric phase as a function of the helix parameters at fixed contour length $L=10 \, \sigma$ is reported in Fig. \ref{fig3}.  We identify the values of the helix radius $r$ and pitch $p$ that give rise to chiral phases, whose handedness with respect to that of the constituent helices is (i) the same, (ii) the opposite, or (iii) mixed, with a chirality inversion from left to right upon increasing the packing fraction. The data displayed in Fig. \ref{fig3} refer to values of the packing fraction $\rho v \leq 0.5$; at higher densities the nematic phase is expected to be metastable with respect to inhomogeneous states \cite{bolhuis}. Fig. \ref{fig3} shows, for $L=10 \, \sigma$, a clear trend with opposite handedness favored by rather elongated helices with $p/\sigma>3$ to $3.5$ We also notice that the cholesteric pitch of particles with small $p$ and $r$, resembling more structured rods with surface roughness rather than proper helices, is very sensitive to small changes in the shape. Nevertheless, our theory unveils this subtle dependence and allows for future, detailed analysis of these complex features. 

\begin{figure}

\includegraphics[scale=0.68]{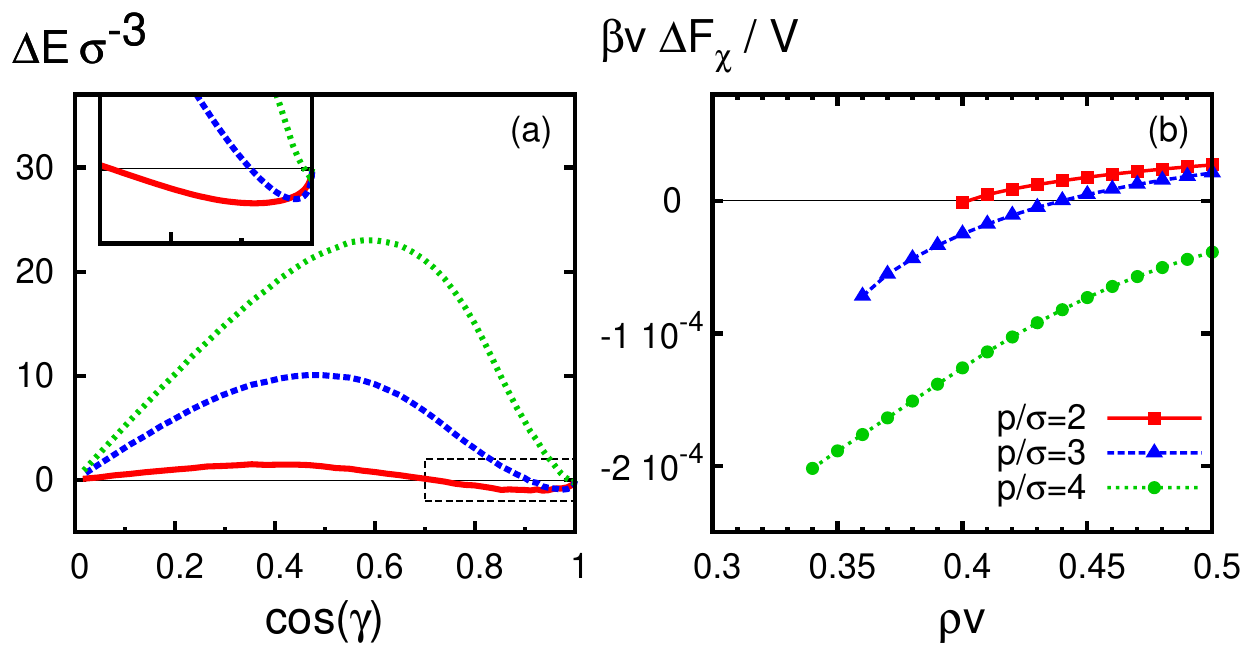}
\caption{\label{fig4} (a) Difference $\Delta E = E_R-E_L$ between the right- and left-handed excluded volume (cf. Eq. (\ref{eq8})) as a function of the cosine of the angle between the main axes $\cos \gamma =\omgvechat \cdot \omgvechat'$ of the same hard helices as in Figs. \ref{fig2}(b)-(c). The dashed rectangle identifies the portion of the plot reported in the inset. (b) Estimated difference in free energy $\Delta F_\chi$ between right- and left-handed chiral ordering as defined in Eq. (\ref{eq9}).}
\end{figure}

Since our model incorporates hard interactions only, the stability of chiral ordering must be due to a gain in excluded-volume entropy with respect to the achiral state. The effect of right-handed chiral ordering consists of favoring right-handed configurations of particle pairs vs. left-handed ones; vice versa for left-handed chiral phases. A pair of rods is in a right-handed (left-handed) configuration if $(\rvec - \rvec')\cdot(\omgvechat \times \omgvechat')>0$ ($<0$). In order to interpret the predictions of the theory, it is crucial to measure the difference in excluded volume between right- and left-handed configurations of pairs. We introduce the right- ($E_R$) and left-handed ($E_L$) excluded volume of two helices as a function of the relative angle between them $\gamma = {\rm acos}(\omgvechat \cdot \omgvechat')$ as

\begin{multline}
E_{\bfrac{R}{L}}(\omgvechat \cdot \omgvechat')=-\int d (\drvec) \int_0^{2 \pi} \frac{d \alpha}{2\pi} \frac{d \alpha'}{2\pi} \times \\
\times f(\drvec, \rrot, \rrot') \, \Theta \bigl(\pm \drvec \cdot (\omgvechat \times \omgvechat') \bigr),
\label{eq8}
\end{multline}
with $\Theta(x)$ the Heaviside step function.
The sum of right- and left-handed excluded volumes in Eq. (\ref{eq8}) gives rise to the usual excluded-volume averaged over the internal azimuthal angles $\alpha$ and $\alpha'$. The difference $\Delta E = E_R-E_L$ between right- and left-handed excluded volume for the three hard-helix models in Figs. \ref{fig2}(b)-(c) is reported in Fig. \ref{fig4}(a). For helices with $p/\sigma=4$ the left-handed excluded volume is smaller than the right-handed one for each value of the angle $\gamma$. Consequently, we expect the resulting chiral phase to be left-handed, as confirmed by Figs. \ref{fig2}(b)-(c). However, for hard helices with $p/\sigma=2$ and $p/\sigma=3$ the same situation arises only when the angle $\gamma$ between the particles is sufficiently larger than zero (i.e., $\cos \gamma $ sufficiently smaller than unity). On the contrary, when $\gamma$ is sufficiently small the excluded volume is minimized by right-handed configurations. This shows why solely the handedness of the particles is not sufficient to determine the handedness of the corresponding chiral phase. Notice that the difference in right- and left-handed excluded volumes in Fig. \ref{fig4}(a) represents a purely geometrical property of the particles. In order to gain further insights, we need to relate such geometric property with the thermodynamics. Mimicking the functional form of the second-virial excess free energy, we estimate the free-energy difference associated to right- and left-handed chiral ordering at density $\rho$ as

\begin{equation}
 \frac{\beta \, \Delta F_{\chi}}{V}= -\frac{\rho^2}{2}\oint d\omgvechat \, d\omgvechat' \psi_0(\nvechat_0 \cdot \omgvechat) \, \psi_0(\nvechat_0 \cdot \omgvechat') \, \Delta E(\omgvechat \cdot \omgvechat'), 
\label{eq9}
\end{equation}
where $\psi_0(\nvechat_0 \cdot \omgvechat)$ is the ODF at density $\rho$ evaluated for simplicity in the achiral limit $q=0$. If at a given density $\rho$ the free energy difference $ \Delta F_\chi$ takes a positive (negative) value, we expect the stable chiral phase to be right-handed (left-handed). We report in Fig. \ref{fig4}(b) the values of $\Delta F_\chi$ for the three hard-helix models considered in Figs. \ref{fig2}(b)-(c). The plots in Fig. \ref{fig4}(b) not only confirm with great accuracy the regimes of stability of the right- and left- handed chiral phases. As manifested by a comparison with Fig. \ref{fig2}(b), they also qualitatively reproduce the density dependence of the chiral wavenumber $q$. 

In conclusion, we developed a DFT for the cholesteric ordering developed by hard chiral rods. Our approach offers a significant improvement over previous attempts to address the problem, since no assumption regarding the length of the pitch, the form of the local ODF or the interactions is required. The only approximation introduced is the Parsons-Lee corrected second-virial truncation of the free energy, which is known to give reliable results when sufficiently thin rods are considered. The algorithm on which our calculations are based allows for the study of arbitrary pair potentials, and can be straightforwardly generalized to account for two-body energetic terms (thermotropic cholesterics). We study the cholesteric ordering of right-handed hard-sphere helices by evaluating the handedness and the pitch of the phase. Our results show that the handedness of the phase depends not only on the details of the interaction, but also on the thermodynamic state. Additionally, by evaluating the separate contribution to the total excluded volume due to right- and left-handed pairs of helices, we uncover relevant insights on the origin of the chiral ordering. The chiral shape of hard helices gives rise to a difference in excluded volume between right- and left-handed pairs of helices. Depending on the local degree of nematic alignment, such a difference can favor, right- or left-handed chiral ordering and, in limit situations, even achiral ordering. Our findings offer a new powerful tool and novel important insights to further advance our understanding of this state of matter.

\smallskip   

This work is financed by a NWO-ECHO grant and is part of the research program of FOM, which is financially supported by NWO.

\end{document}